# The Hill Stability of Triple Minor Planets in the Solar System[*]

Xiaodong Liu,[1] Hexi Baoyin,[1][★] Nikolaos Georgakarakos,[2] John Richard Donnison[3] and Xingrui Ma[1]

[1]*School of Aerospace, Tsinghua University, Beijing 100084, China*

[2]*128 V. Olgas str., Thessaloniki 54645, Greece*

[3]*Astronomy Unit, School of Physics and Astronomy, Queen Mary, University of London, Mile End Road, London E1 4NS, England*

## ABSTRACT

The triple asteroids and triple Kuiper belt objects (collectively called the triple minor planets) in the Solar system are of particular interest to the scientific community since the discovery of the first triple asteroid system in 2004. In this paper, the Hill stability of the nine known triple minor planets in the Solar system is investigated. Seven of the systems are of large size ratio, i.e. they consist of a larger primary and two moonlets, while the other two systems have components of comparable size. Each case is treated separately. For the triple minor planets that have large size ratio, the sufficient condition for Hill stability is expressed in closed form. This is not possible for the systems with comparable size components, for which the

---

Hill stability is assessed by a combination of analytical and numerical means. It is found that all the known triple minor planets are Hill stable, except 3749 Balam, for which the incomplete orbital parameters make the Hill stability of the system uncertain. This suggests that there might be more such stable triple minor planets in the Solar system yet to be observed. It is also shown that the Hill stability regions increase as the mutual inclination between the inner orbit and outer orbit decreases, the semimajor axis ratio of the inner orbit with respect to the outer orbit decreases, and the mass ratio of the outer satellite with respect to the inner satellite increases. This study therefore provides useful information about dynamical properties of the triple minor planets in the Solar system.

## 1 INTRODUCTION

Since the first discovery of the binary asteroid system Ida-Dactyl in 1993 (Chapman et al. 1995; Belton et al. 1995, 1996), the investigation of multiple minor planets has attracted great attention (Richardson & Walsh 2006; Johnston 2012). The system 87 Sylvia is the first known triple asteroid system to be observed, where its second moon was discovered in 2004 (Marchis et al. 2005). At the time of writing this paper, nine triple minor planets have been identified in the Solar system, namely, 45 Eugenia (Marchis et al. 2010; Beauvalet et al. 2011), 87 Sylvia (Marchis et al. 2005;

Fang, Margot & Rojo 2012), 93 Minerva (Marchis et al. 2011), 216 Kleopatra (Descamps et al. 2011), 1994 CC (Fang et al. 2011; Brozović et al. 2011), 2001 SN263 (Fang et al. 2011), 136108 Haumea (Ragozzine & Brown 2009), 1999 TC$_{36}$ (Benecchi et al. 2010), and 3749 Balam (Marchis et al. 2008; Marchis et al. 2012). However, the orbital parameters of 3749 Balam are still incomplete.

In this study, the Hill stability of the nine known triple minor planets in the Solar system is examined. The notion of Hill stability dates back to many years ago, and its concept was introduced by Hill (1878) when he was studying the restricted three-body problem. The history of Hill stability was well reviewed in (Georgakarakos 2008, Li, Fu & Sun 2010, Donnison 2011), which the reader is referred to for details. It is well known that the regions of possible motions for the restricted three-body problem depend on the Jacobi constant. While for the general three-body problem, as an extension, the parameter $c^2E$ determines the regions of possible motions (e.g. Marchal & Saari 1975; Zare 1976), where $c$ is the total angular momentum and $E$ is the total energy of the system. To judge whether a triple system is Hill stable, the actual value of $c^2E$ for the real system is compared with the critical value of the same parameter for the collinear equilibrium configurations (Zare 1976; Szebehely & Zare 1977). If the actual value of $c^2E$ is less than the critical value, the zero-velocity surface is closed, so the system is Hill stable and no exchange of the components will happen. If the actual value of $c^2E$ is greater than the critical value, the zero-velocity surface opens out, so the system is not Hill stable and the exchange of the components is possible. Note that the Hill stability cannot provide information about the possibility of escape of the third body from the triple system. In previous studies, numerical simulations were used to investigate the orbital stability of the specific triple asteroids 87 Sylvia (Winter et al 2009; Frouard and Compère 2012). Numerical integrations were also performed to analyze the stability regions around the triple asteroids 2001 SN263 (Araujo et al. 2012).

The theory of the Hill stability was first applied to planar cases including triple stars, pairs of planets orbiting the Sun, Planet-Satellite-Sun systems and Planet-Planet-Star systems (Szebehely & Zare 1977; Walker & Roy 1981; Marchal & Bozis 1982; Walker 1983; Donnison & Williams 1983; Valsecchi, Carusi & Roy 1984; Donnison 1984, 1988; Gladman 1993; Kiseleva, Eggleton & Orlov 1994; Donnison & Mikulskis 1992, 1994, 1995). More recently, the theory of Hill stability was applied to the situation where the third body is inclined to the binary. The applications include triple star systems (Donnison 2010a), extrasolar planetary systems (Veras & Armitage 2004; Donnison 2006; Donnison 2009; Donnison 2010a,b), Kuiper Belt binary systems (Donnison 2008; Donnison 2010a; Li et al. 2010; Donnison 2011), and binary asteroids (Donnison 2011).

To our knowledge, this paper may be the first to examine the Hill stability of the triple minor planets in the Solar system. The paper is structured as follows: in the next section, the dynamical model of the system is described, and the exact expressions for the total energy and the total angular momentum of the system are derived. In Section 3, for large size ratio triple minor planets, the actual value and critical value of $c^2E$ are expanded in terms of the mass ratio, and the sufficient condition for Hill stability is expressed in closed form. For triple minor planets of comparable size, the extreme value of $c^2E$ is computed and compared with its critical value calculated numerically at the collinear equilibrium configurations in Section 4. Finally, our discussion and conclusions are presented.

## 2 THE DYNAMICAL MODEL OF THE SYSTEM

The triple minor planets consist of three components: the primary with mass $m_1$, and two satellites with masses $m_2$ and $m_3$, respectively. The hierarchical arrangement where $m_1$ and $m_2$ form a binary pair and $m_3$ is an external component is adopted (Walker & Roy, 1981). It is assumed that $m_2$ is in orbit about $m_1$ with orbital elements ($a_2$, $e_2$, $i_2$, $\Omega_2$, $\omega_2$, $f_2$), and $m_3$ is in orbit about $\mu = m_1 + m_2$, located at the barycentre of the binary pair ($m_1$, $m_2$), with orbital elements ($a_3$, $e_3$, $i_3$, $\Omega_3$, $\omega_3$, $f_3$). The orbital elements used above are defined as follows: $a$ is the semimajor axis, $e$ is the eccentricity, $i$ is the inclination, $\Omega$ is the longitude of the ascending node, $\omega$ is the argument of pericentre, and $f$ is the true anomaly. The orbital elements with subscripts "2" and "3" denote that they are orbital parameters of $m_2$'s orbit about $m_1$ and $m_3$'s orbit about $\mu$, respectively. Table 1 lists some physical and orbital parameters for the nine known triple minor planets, where $\lambda$ is the mass ratio of $m_3$ and $m_2$, i.e., $\lambda = m_3 / m_2$; $\alpha$ is the semimajor axis ratio, i.e., $\alpha = a_2 / a_3$, and $I$ is the mutual inclination of $m_2$'s orbit about $m_1$ with respect to $m_3$'s orbit about $\mu$.

**Table 1.** Physical and orbital parameters for the nine known triple minor planets.

| System | $m_1$ (kg) | $m_2$ (kg) | $m_3$ (kg) | $a_2$ (km) | $a_3$ (km) | $e_2$ | $e_3$ | $\lambda$ | $\alpha$ | $I$ (°) | $a_2/R_{H2}$ | $a_3/R_{H3}$ | Bifurcation |
|---|---|---|---|---|---|---|---|---|---|---|---|---|---|
| Eugenia[1,2] | $5.63 \times 10^{18}$ | $2.51 \times 10^{14}$ | $2.51 \times 10^{14}$ | 610.79 | 1164.42 | 0.078 | 0.0004 | 1.00 | 0.52 | 20.7 | 0.015 | 0.029 | Primary |
| Sylvia[3] | $1.484 \times 10^{19}$ | $7.333 \times 10^{14}$ | $9.319 \times 10^{14}$ | 706.5 | 1357 | 0.02721 | 0.005566 | 1.27 | 0.52 | 0.56 | 0.010 | 0.019 | Primary |
| Minerva[4] | $3.60 \times 10^{18}$ | $2.12 \times 10^{13}$ | $5.03 \times 10^{14}$ | 385 | 651 | 0.005 | 0 | 2.37 | 0.59 | 5.4 | 0.011 | 0.019 | Primary |
| Kleopatra[5] | $4.64 \times 10^{18}$ | $6.49 \times 10^{14}$ | $1.39 \times 10^{15}$ | 454 | 678 | 0 | 0 | 2.14 | 0.67 | 5.0 | 0.012 | 0.018 | Primary |
| Haumea[6] | $4.006 \times 10^{21}$ | $1.79 \times 10^{18}$ | $1.79 \times 10^{19}$ | 25657 | 49880 | 0.249 | 0.0513 | 10.00 | 0.51 | 13.4 | 0.005 | 0.009 | Primary |
| CC[7] | $25.935 \times 10^{10}$ | $0.580 \times 10^{10}$ | $0.091 \times 10^{10}$ | 1.729 | 6.130 | 0.002 | 0.192 | 0.16 | 0.28 | 15.7 | 0.020 | 0.071 | Secondary |
| SN263[7] | $917.466 \times 10^{10}$ | $9.773 \times 10^{10}$ | $24.039 \times 10^{10}$ | 3.804 | 16.633 | 0.016 | 0.015 | 2.46 | 0.23 | 13.9 | 0.011 | 0.048 | Primary |
| TC$_{36}$[8] | $6.002 \times 10^{18}$ | $6.002 \times 10^{18}$ | $0.746 \times 10^{18}$ | 867 | 7411 | 0.101 | 0.2949 | 0.12 | 0.12 | 10.7 | 0.001 | 0.010 | Secondary |
| Balam[9,10,11] | $1.01 \times 10^{14}$ | $6.24 \times 10^{12}$ | $7.41 \times 10^{12}$ | 20 | 203.4 | | 0.573 | 1.19 | 0.10 | | 0.023 | 0.227 | Primary |

**Reference.** 1. Marchis et al. (2010); 2. Beauvalet et al. (2011); 3. Fang et al. (2012); 4. Marchis et al. (2011); 5. Descamps et al. (2011); 6. Ragozzine and Brown (2009); 7. Fang et al. (2011); 8. Benecchi et al. (2010); 9. Vachier, Berthier & Marchis (2012); 10. Marchis et al. (2012); 11. Johnston (2012).

The mutual Hill radii $R_{H2}$ and $R_{H3}$ of the pairs ($m_1$, $m_2$) and ($\mu$, $m_3$) are defined in equations (1) and (2) respectively (Donnison & Williams 1975; Richardson & Walsh 2006):

$$R_{H2} = a_\odot \left[\frac{m_1 + m_2}{3m_\odot}\right]^{1/3}, \tag{1}$$

$$R_{H3} = a_\odot \left[\frac{\mu + m_3}{3m_\odot}\right]^{1/3}, \tag{2}$$

where $a_\odot$ is the semimajor axis of the binary pair barycentre's orbit about the Sun; and $m_\odot$ is the mass of the Sun. Within the mutual Hill radius, the pair's mutual gravity dominates the attraction of the components compared to the solar gravity. The values of $a_2/R_{H2}$ and $a_3/R_{H3}$ are listed in columns (12) and (13) of Table 1, respectively. It can be seen that the mutual Hill radii for all the nine known triple minor planets are all much greater than the companions' semimajor axis. Thus, the effect of the solar tidal force is ignored in this paper, and the triple minor planets can be treated as an isolated system, which means that only the mutual gravity of the hierarchical triple systems is considered.

For the triple system, the exact expression of the total energy instead of the two-body approximation is used (Walker & Roy 1981; Valsecchi et al. 1984):

$$E = -\frac{Gm_1 m_2}{2a_2} - \frac{Gm_1 m_3}{r_{31}} - \frac{Gm_2 m_3}{r_{23}} + \frac{Gm_3 (m_1 + m_2)}{2a_3 (1-e_3^2)}\left(1 + 2e_3 \cos f_3 + e_3^2\right), \tag{3}$$

where $r_{23}$ is the distance between $m_2$ and $m_3$, and $r_{31}$ is the distance between $m_1$ and $m_3$:

$$r_{23} = \left[\rho_3^2 + \frac{m_1^2}{(m_1 + m_2)^2}\rho_2^2 - \frac{2m_1}{m_1 + m_2}\rho_2 \rho_3 \cos\theta\right]^{1/2}, \tag{4}$$

$$r_{31} = \left[ \rho_3^2 + \frac{m_2^2}{(m_1 + m_2)^2} \rho_2^2 + \frac{2m_2}{m_1 + m_2} \rho_2 \rho_3 \cos\theta \right]^{1/2}, \tag{5}$$

where $\theta$ is the phase angle of the system given by

$$\cos\theta = (\boldsymbol{\rho}_2 \cdot \boldsymbol{\rho}_3)/(\rho_2 \rho_3); \tag{6}$$

$\boldsymbol{\rho}_2$ and $\boldsymbol{\rho}_3$ are the orbital radius vector of $m_2$ with respect to $m_1$, and the orbital radius vector of $m_3$ with respect to $\mu$, respectively; $\rho_2 = |\boldsymbol{\rho}_2|$, and $\rho_3 = |\boldsymbol{\rho}_3|$; and $G$ is the universal gravitational constant. The exact expression of the total angular momentum of the system is given by

$$\boldsymbol{c} = \frac{m_1 m_2}{(m_1 + m_2)^{1/2}} \left[ Ga_2(1-e_2^2) \right]^{1/2} \boldsymbol{h}_2 + \frac{m_3(m_1+m_2)}{(m_1+m_2+m_3)^{1/2}} \left[ Ga_3(1-e_3^2) \right]^{1/2} \boldsymbol{h}_3, \tag{7}$$

where $\boldsymbol{h}_2$ is the unit vector along the orbital angular momentum of $m_2$ with respect to $m_1$; and $\boldsymbol{h}_3$ is the unit vector along the orbital angular momentum of $m_3$ with respect to $\mu$.

It is known that the parameter $c^2E$ determines the regions of possible motion of the system. Since $G$ is constant, the parameter $S$ is used as a substitute for $c^2E$,

$$S = c^2 |E| / G^2. \tag{8}$$

The value of $S_{ac}$ for the actual system is calculated by

$$S_{ac} = \left\{ c^2 |E| / G^2 \right\}_{ac}. \tag{9}$$

On the substitution of equations (3) and (7) into equation (9), the full expression of $S_{ac}$ is written as

$$S_{ac} = \frac{k\mu^3 m_3^3}{2M}(1-e_3^2) + \frac{km_1^2 m_2^2 m_3}{2}(1-e_2^2)\alpha - \frac{\mu^2 m_2 m_3^3}{M}\frac{a_3}{r_{23}}(1-e_3^2)$$

$$-\frac{m_1 m_2 \mu^2 m_3^2}{2M}\frac{1-e_3^2}{\alpha} - \frac{m_1^3 m_2^3}{2\mu}(1-e_2^2) - m_1^2 m_2^2 m_3 \cos I \left[\frac{\mu(1-e_2^2)(1-e_3^2)}{M\alpha}\right]^{1/2}$$

$$+km_1 m_2 m_3^2 \cos I \left[\frac{\mu^3(1-e_2^2)(1-e_3^2)\alpha}{M}\right]^{1/2} - \frac{m_1^2 m_3^3 m_3}{\mu}\frac{a_2}{r_{23}}(1-e_2^2) \quad (10)$$

$$-\frac{2m_1 m_2^2 m_3^2}{r_{23}}\cos I \left[\frac{\mu a_2 a_3 (1-e_2^2)(1-e_3^2)}{M}\right]^{1/2} - \frac{\mu^2 m_1 m_3^3}{M}\frac{a_3}{r_{31}}(1-e_3^2)$$

$$-\frac{m_1^3 m_2^2 m_3}{\mu}\frac{a_2}{r_{31}}(1-e_2^2) - \frac{2m_1^2 m_2 m_3^2}{r_{31}}\cos I \left[\frac{\mu a_2 a_3 (1-e_2^2)(1-e_3^2)}{M}\right]^{1/2},$$

where $k = (1 + 2e_3 \cos f_3 + e_3^2)/(1-e_3^2)$. The parameter $M = m_1 + m_2 + m_3$ is the total mass of the system. The mutual inclination $I$ is given by

$$\cos I = \mathbf{h}_2 \cdot \mathbf{h}_3 = \sin i_2 \sin i_3 \cos(\Omega_2 - \Omega_3) + \cos i_2 \cos i_3, \quad (11)$$

or

$$\cos I = \cos \beta_{h_2} \cos \beta_{h_3} \cos(\lambda_{h_2} - \lambda_{h_3}) + \sin \beta_{h_2} \sin \beta_{h_3}, \quad (12)$$

where $\lambda_{h_2}$ and $\beta_{h_2}$ are the ecliptic longitude and latitude of orbital pole orientation of the pair ($m_1$, $m_2$), respectively; $\lambda_{h_3}$ and $\beta_{h_3}$ are the ecliptic longitude and latitude of orbital pole orientation of the pair ($\mu$, $m_3$), respectively.

As seen from column (11) of Table 1, for the triple asteroids 87 Sylvia (Fang et al. 2012), 93 Minerva (Marchis et al. 2011), and 216 Kleopatra (Descamps et al. 2011), $m_2$'s orbit about $m_1$ and $m_3$'s orbit about $\mu$ are almost coplanar, i.e., the mutual inclination $I$ is small. For 2001 SN263 (Fang et al. 2011), 1994 CC (Fang et al. 2011; Brozović et al. 2011), 45 Eugenia (Marchis et al. 2010; Beauvalet et al. 2011), 136108 Haumea (Ragozzine & Brown 2009), and 1999 TC$_{36}$ (Benecchi et al. 2010), the mutual inclination $I$ between the inner orbit and the outer orbit is significant. However, this paper does not separate these two cases.

The critical value of the parameter $S$, i.e., $S_{cr}$, is determined for the collinear equilibrium configurations (Zare 1976; Szebehely & Zare 1977). Assuming that ratio of the distances is taken as $|m_1 m_2|:|m_2 m_3| = 1:x$, it is known that $x$ satisfies the following quintic equation (Roy 1978, p124)

$$(m_1 + m_2)x^5 + (3m_1 + 2m_2)x^4 + (3m_1 + m_2)x^3 - (3m_3 + m_2)x^2 \\ - (3m_3 + 2m_2)x - (m_3 + m_2) = 0, \tag{13}$$

and $x$ is the unique positive root of equation (13).

It is known that there are three possible bifurcation points in the collinear equilibrium configurations: the primary, secondary and tertiary bifurcation points (Zare 1977; Szebehely & Zare 1977; Walker & Roy 1981), the values of $S$ for which are denoted as $S_1$, $S_2$, and $S_3$, respectively. The primary bifurcation requires the smallest mass to be positioned in the centre, the secondary and tertiary bifurcation require the intermediate mass and the largest mass to be so positioned, respectively (Walker & Roy 1981). The values of $S_1$, $S_2$, and $S_3$ satisfy $S_1 \geq S_2 \geq S_3$ (Walker & Roy 1981). If $S > S_1$, no exchange will occur between bodies. If $S_2 < S < S_1$, then also there is no exchange between bodies if the smallest mass is the external body (Zare 1977). As seen from columns (9) and (14) of Table 1, for 87 Sylvia (Fang et al. 2012), 93 Minerva (Marchis et al. 2011), 216 Kleopatra (Descamps et al. 2011), 45 Eugenia (Marchis et al. 2010), 2001 SN263 (Fang et al. 2011), 136108 Haumea (Ragozzine & Brown 2009), and 3749 Balam (Marchis et al. 2012), $\lambda > 1$, i.e., $m_2$ is the smallest mass, so the primary bifurcation is sought, which corresponds to the order of masses $m_1$-$m_2$-$m_3$. Note that for 45 Eugenia (Marchis et al. 2010), $m_2$ is slightly smaller than $m_3$. For 1994 CC (Fang et al. 2011), and 1999 TC$_{36}$ (Benecchi et al. 2010), $\lambda < 1$, i.e., the smallest mass $m_3$ is the external body, so the secondary bifurcation determines the critical $S$, which also corresponds to the order of masses $m_1$-$m_2$-$m_3$. Therefore, for either the primary or the secondary bifurcation described above, the required order of masses in our problem is $m_1$-$m_2$-$m_3$ (Zare 1977; Walker & Roy 1981). In this order, the critical value $S_{cr}$ is calculated by (Zare 1976)

$$S_{cr} = \{c^2|E|/G^2\}_{cr} = f^2(x)g(x)/2M, \tag{14}$$

where

$$f(x) = m_1 m_2 + m_1 m_3/(1+x) + m_2 m_3/x, \tag{15}$$

and

$$g(x) = m_1 m_2 + m_1 m_3 (1+x)^2 + m_2 m_3 x^2. \tag{16}$$

The expression of $S_{cr}$ is presented in Donnison & Williams (1983), and it is reproduced here as

$$\begin{aligned}
S_{cr} = & \frac{m_1^3 m_2^3 + m_1^3 m_3^3 + m_2^3 m_3^3}{2M} + \frac{m_1^2 m_2^2 m_3^2}{M}\left[\frac{1+x^3+(1+x)^3}{x(1+x)}\right] + \frac{m_1^3 m_2^2 m_3}{2M}\left[\frac{2+(1+x)^3}{1+x}\right] \\
& + \frac{m_1^2 m_2^3 m_3}{2M}\left(\frac{2+x^3}{x}\right) + \frac{m_1^3 m_2 m_3^2}{2M}\left[\frac{1+2(1+x)^3}{(1+x)^2}\right] + \frac{m_1 m_2^3 m_3^2}{2M}\left(\frac{1+2x^3}{x^2}\right) \\
& + \frac{m_1^2 m_2 m_3^3}{2M}\left[\frac{2(1+x)^3+x^3}{x(1+x)^2}\right] + \frac{m_1 m_2^2 m_3^3}{2M}\left[\frac{(1+x)^3+2x^3}{x^2(1+x)}\right].
\end{aligned} \tag{17}$$

## 3 LARGE SIZE RATIO TRIPLE MINOR PLANETS

Among the nine known triple minor planets, 45 Eugenia (Marchis et al. 2010), 87 Sylvia (Fang et al. 2012), 93 Minerva (Marchis et al. 2011), 216 Kleopatra (Descamps et al. 2011), 1994 CC (Fang et al. 2011), 2001 SN263 (Fang et al. 2011), and 136108 Haumea (Ragozzine & Brown 2009) are large size ratio systems; while 1999 TC$_{36}$ (Benecchi et al. 2010) and 3749 Balam (Marchis et al. 2012) have components which are of comparable size. In this section, the large size ratio systems are considered.

For the large size ratio triple minor planets, $m_1 \gg m_2 + m_3$. Thus, an approximate solution of equation (13) is obtained (Walker, Emslie & Roy 1980)

$$x_0 \approx \left(\frac{m_2 + m_3}{3m_1}\right)^{1/3}. \tag{18}$$

Since $x_0$ is small, the critical value $S_{cr}$ can be expanded in terms of $x_0$,

$$S_{cr} = \chi\left[(1+\lambda)^3 + O(x_0^2)\right], \tag{19}$$

where $\chi = m_1^2 m_2^3 / 2$.

Similarly, the actual value $S_{ac}$ is expanded in terms of $x_0$,

$$\begin{aligned}S_{ac} = \chi\Big[&\lambda^3(1-e_3^2) + \lambda(1-e_2^2)\alpha + 2\lambda^2 \cos I (1-e_2^2)^{1/2}(1-e_3^2)^{1/2}\alpha^{1/2} \\ &+\lambda^2(1-e_3^2)\alpha^{-1} + (1-e_2^2) + 2\lambda \cos I (1-e_2^2)^{1/2}(1-e_3^2)^{1/2}\alpha^{-1/2} + O(x_0^3)\Big].\end{aligned} \tag{20}$$

The sufficient condition for Hill stability is expressed in terms of $\lambda$ as

$$S_{ac} - S_{cr} = A\lambda^3 + B\lambda^2 + C\lambda + D > 0, \tag{21}$$

where

$$\begin{aligned}&A = -e_3^2, \quad B = 2\cos I (1-e_2^2)^{1/2}(1-e_3^2)^{1/2}\alpha^{1/2} + (1-e_3^2)\alpha^{-1} - 3, \\ &C = (1-e_2^2)\alpha + 2\cos I (1-e_2^2)^{1/2}(1-e_3^2)^{1/2}\alpha^{-1/2} - 3, \quad D = -e_2^2.\end{aligned} \tag{22}$$

If $e_3 \neq 0$, i.e., $A \neq 0$, the discriminant of the cubic equation (21) can be written as

$$\Delta_3 = \left(\frac{q}{2}\right)^2 + \left(\frac{p}{3}\right)^3 \tag{23}$$

where

$$p = -\frac{B^2}{3A^2} + \frac{C}{A}, \quad q = \frac{2B^3}{27A^3} - \frac{BC}{3A^2} + \frac{D}{A}. \tag{24}$$

If $\Delta_3 \leq 0$, equation (21) possesses three real roots:

$$\lambda_j = -\frac{B}{3A} + 2\sqrt{-\frac{p}{3}} \cos\left[\frac{1}{3}\arccos\left(\frac{3q}{2p}\sqrt{\frac{-3}{p}}\right) - j\frac{2\pi}{3}\right], \text{ for } j = 0, 1, 2, \qquad (25)$$

We denote by $\lambda_{3max}$, $\lambda_{3med}$, and $\lambda_{3min}$ the maximum, intermediate, and minimum real roots of equation (21), respectively. Because $A < 0$, the solution of equation (21) should satisfy

$$\lambda < \lambda_{3min}, \text{ or } \lambda_{3med} < \lambda < \lambda_{3max}. \qquad (26)$$

If $\Delta_3 > 0$, equation (21) possesses only one real root:

$$\lambda_{sig} = -\frac{B}{3A} + \left(-q/2 + \Delta_3^{3/2}\right)^{1/3} + \left(-q/2 - \Delta_3^{3/2}\right)^{1/3}, \qquad (27)$$

so the value of $\lambda$ should satisfy

$$\lambda > \lambda_{sig}. \qquad (28)$$

If $e_3 = 0$, i.e., $A = 0$, equation (21) is a quadratic equation, and the case is simpler. The discriminant of equation (21) can be written as

$$\Delta_2 = C^2 - 4BD. \qquad (29)$$

If $\Delta_2 > 0$, we denote by $\lambda_{2max}$ and $\lambda_{2min}$ the maximum and minimum roots of equation (21), respectively. Therefore, to ensure Hill stability, the following conditions given by equations (30-32) should be met:

If $B > 0$,

$$\Delta_2 \geq 0, \text{ and } \lambda_{2min} < \lambda < \lambda_{2max}; \qquad (30)$$

If $B < 0$,

$$\Delta_2 < 0 \text{ and no solution for } \lambda, \text{ or } \Delta_2 \geq 0 \text{ and } \lambda > \lambda_{2max}; \qquad (31)$$

If $B = 0$,

$$C > 0 \text{ and } \lambda > -B/C, \text{ or } C < 0 \text{ and } \lambda < -B/C; \qquad (32)$$

However, there are too many variables intricately coupled with each other in equation

(21), so it is difficult to determine the relation of the variables to ensure the Hill stability. In order to simplify the analysis, the values of $e_2$ and $e_3$ are kept constant as their actual observed values. Figs. 1(a)-1(g) show the critical mass ratio $\lambda$ versus the semimajor axis ratio $\alpha$ and the mutual inclination $I$ for the specific triple minor planets' $e_2$ and $e_3$ using equations (21-32). The positions of the actual large size ratio triple minor planets in the parameter space are also marked with red dots in Figs. 1(a)-1(g), using the parameters listed in columns (9-11) of Table 1. The systems below the critical Hill stability surface are Hill stable, where the mass exchange of the components is precluded. The systems above the surface could be Hill unstable, where the exchange of the components is possible. It can be seen that the triple minor planets 45 Eugenia, 87 Sylvia, 93 Minerva, 216 Kleopatra, 136108 Haumea, 2001 SN263, and 1994 CC are all Hill stable. Thus, mass exchange or collision of the components is not possible. Note that the Hill stability of a system is not the same as its orbital stability, where the latter only allows small changes in the orbital elements. In the previous studies, numerical simulations were used to analyze the orbital stability of the triple asteroids 87 Sylvia considering the non-sphericity of Sylvia, the mutual perturbation of the moonlets, and the solar perturbation, etc. (Winter et al 2009; Frouard and Compère 2012), and it was found that the triple system 87 Sylvia is orbitally stable.

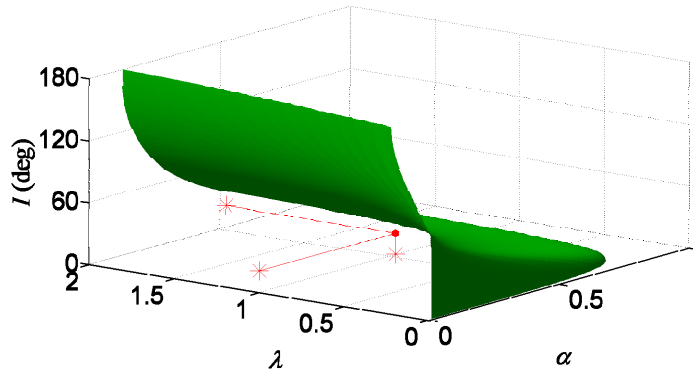
(**a**) 45 Eugenia

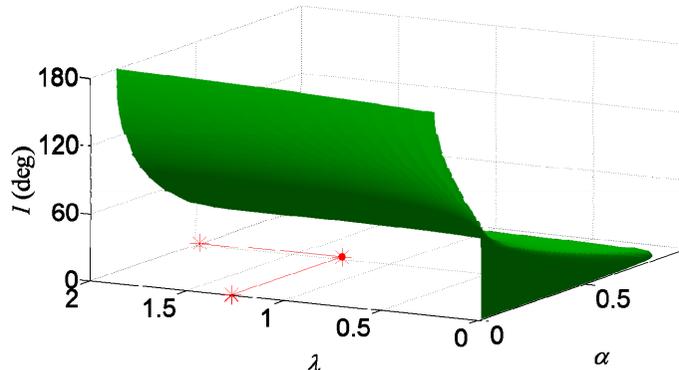
(**b**) 87 Sylvia

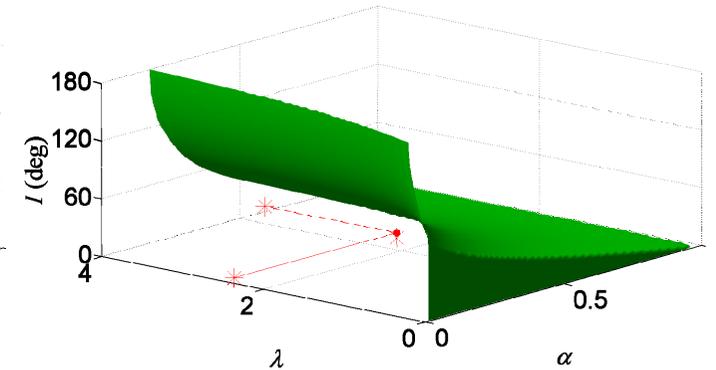
(**c**) 93 Minerva

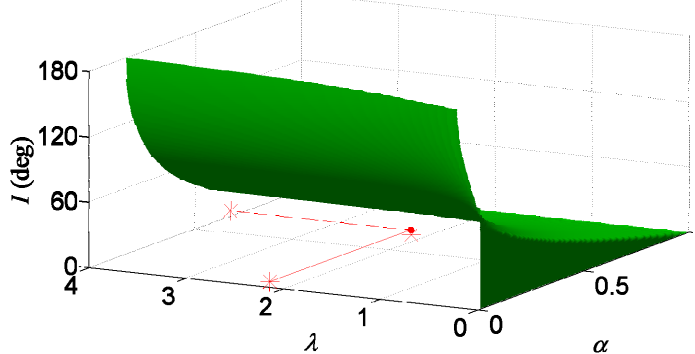
(**d**) 216 Kleopatra

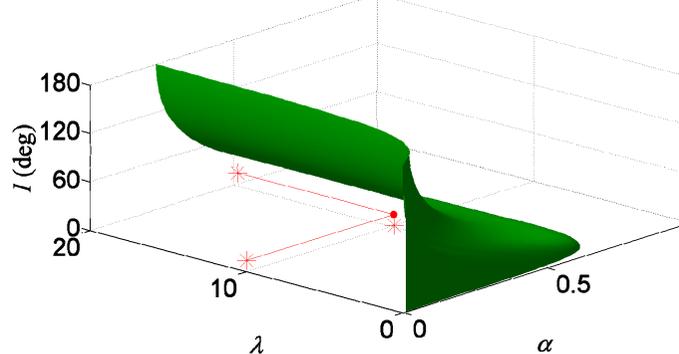
(**e**) 136108 Haumea

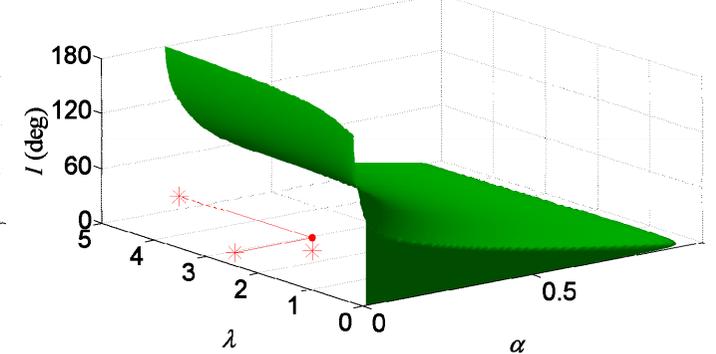
(**f**) 2001 SN263

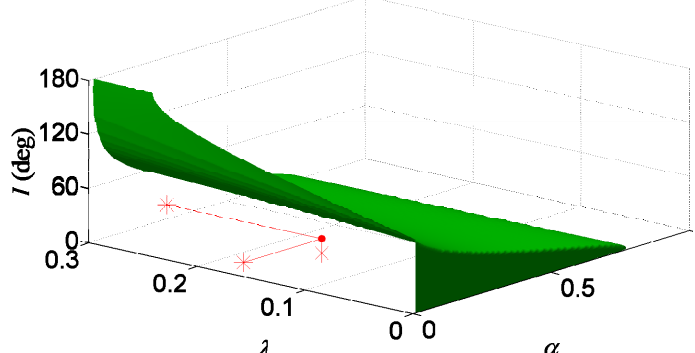
(**g**) 1994 CC

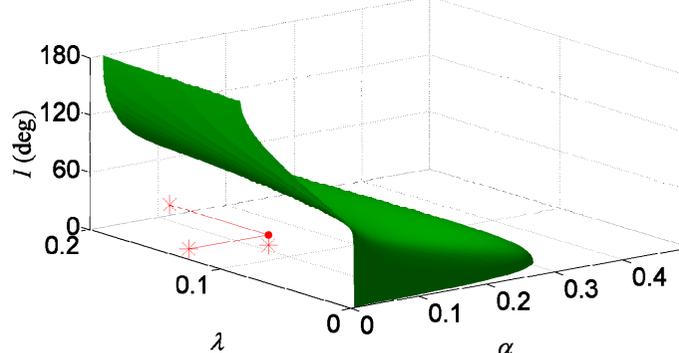
(**h**) 1999 TC$_{36}$

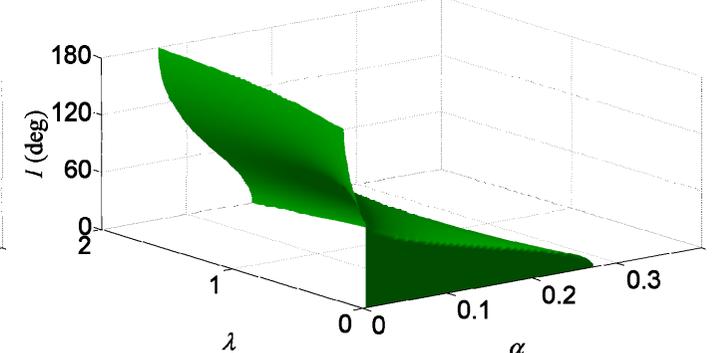
(**i**) 3749 Balam

**Figure 1.** The critical mass ratio $\lambda$ versus the semimajor axis ratio $\alpha$ and mutual inclination $I$ with the constant triple minor planets' $e_2$ and $e_3$. The critical mass ratio $\lambda$ is calculated using equations (21-32). Systems below the critical surface are Hill stable. Using the parameters listed in columns (9-11) of Table 1, the actual triple cases are marked with red dots. The red dots' projections onto the coordinate planes $\alpha = 0$, $I = 0$, and the $\lambda$ = constant plane (the value of constant is set to the upper boundary of the coordinate $\lambda$ in the frame) are marked with red asterisk. Connections between the dots and their projections are marked with in red lines.

Previous studies about the binary systems orbiting the Sun showed that the mutual inclination relative to the Sun has little effect on Hill stability (Li et al. 2010; Donnison 2011). However, in this study, it is found that the mutual inclination $I$ within the triple minor planets themselves is an important factor for Hill stability. As seen from Figs. 1(a)-1(g), the Hill stability regions broaden as the mutual inclination $I$ decreases, the semimajor axis ratio $\alpha$ of the inner pair ($m_1$, $m_2$) with respect to the outer pair ($\mu$, $m_3$) decreases, and the mass ratio of the outer satellite with respect to the inner satellite $\lambda$ increases.

## 4 TRIPLE MINOR PLANETS WITH COMPONENTS OF COMPARABLE SIZE

For the triple minor planets 1999 TC$_{36}$ (Benecchi et al. 2010) and 3749 Balam (Marchis et al. 2012), the triple system consists of a primary that is and composed of two similar-sized components, and a satellite. These triple minor planets are of comparable size, i.e., there is not a dominant mass which is much larger than the other two in the system, so the expressions for $S_{cr}$ and $S_{ac}$ cannot be expanded in terms of the mass ratio. The sufficient condition for Hill stability cannot therefore be expressed in closed form. Thus, the values of $S_{cr}$ and $S_{ac}$ can only be calculated numerically and separately. The original forms of $S_{cr}$ and $S_{ac}$ should be used.

It is known that equation (13) possesses only one positive root (Roy 1978, p124), i.e., $x$. The value of $x$ has to be found numerically from equation (13). Substituting $x$ into equation (17), the critical value $S_{cr}$ can then be obtained.

While for the actual value $S_{ac}$, the extremum conditions are applied. Based on Walker & Roy (1981), the value of $S_{ac}$ reaches the minimum with respect to $\theta$ for the coplanar case when

$$\theta = \theta_m, \tag{33}$$

where

$$\cos\theta_m = \frac{\rho_2(m_1 - m_2)}{2\rho_3(m_1 + m_2)}, \tag{34}$$

which is also applicable to the inclined case. It is also found that when $\theta = \theta_m$, the value of $S_{ac}$ reaches the minimum $\{S_{ac}\}_{min}$ when $f_2 = \pm\pi$, and $f_3 = 0$ with respect to $f_2$ and $f_3$, respectively. In this condition, the exact expression of $\{S_{ac}\}_{min}$ can be written as

$$\{S_{ac}\}_{min} = \left[-\frac{m_1 m_2}{2a_2} + \frac{m_3\mu(1+e_3)}{2a_3(1-e_3)} - \frac{m_3\mu}{r_m}\right] \cdot \left\{\frac{m_1^2 m_2^2}{\mu}a_2(1-e_2^2) + \frac{m_3^2\mu^2}{M}a_3(1-e_3^2) + \frac{2m_1 m_2 m_3}{M^{1/2}}\left[\mu a_2 a_3(1-e_2^2)(1-e_3^2)\right]^{1/2}\cos I\right\}, \tag{35}$$

where

$$r_m = \left[a_3^2(1+e_3)^2 + \frac{m_1 m_2}{\mu^2}a_2^2(1-e_2)^2\right]^{1/2}. \tag{36}$$

After some simple algebraic operations, equation (35) can be rewritten in terms of $\alpha$

as

$$\{S_{ac}\}_{\min} = \left[-\frac{m_1 m_2}{2} + \frac{\alpha m_3 \mu (1+e_3)}{2(1-e_3)} - \frac{m_3 \mu}{\bar{r}_m}\right] \cdot \left\{\frac{m_1^2 m_2^2}{\mu}(1-e_2^2) + \frac{m_3^2 \mu^2 (1-e_3^2)}{\alpha M}\right.$$
$$\left. + \frac{2 m_1 m_2 m_3}{(\alpha M)^{1/2}} \left[\mu(1-e_2^2)(1-e_3^2)\right]^{1/2} \cos I \right\}, \qquad (37)$$

where

$$\bar{r}_m = \left[\frac{(1+e_3)^2}{\alpha^2} + \frac{m_1 m_2}{\mu^2}(1-e_2)^2\right]^{1/2}. \qquad (38)$$

On the substitution of $\lambda = m_3/m_2$ into equations (17) and (37), the sufficient condition for Hill stability can be written as:

$$\{S_{ac}\}_{\min} - S_{cr} = f(\lambda, \alpha, I) > 0. \qquad (39)$$

Fig. 1(h) shows the critical mass ratio $\lambda$ versus the semimajor axis ratio $\alpha$ and mutual inclination $I$ with the 1999 TC$_{36}$ system's $e_2$ and $e_3$ using equations (13), (37), and (39). The position of 1999 TC$_{36}$ in the parameter space is also marked with a red dot in Fig. 1(h), using the parameters listed in columns (9-11) of Table 1. The Hill stability is assured for the systems below the critical Hill stability surface. It can be seen that the triple system 1999 TC$_{36}$ is Hill stable.

At the time of writing, the orbital parameters of 3749 Balam are still incomplete. It is assumed that $e_2 = 0$. The critical Hill stability surface in the $\lambda$, $\alpha$, and $I$ space is shown in Fig. 1(i). If the values for the system 3749 Balam are below the critical surface, it is Hill stable; otherwise, it is Hill unstable. For the triple minor planets 1999 TC$_{36}$ and 3749 Balam, the Hill stability regions also broaden as the mutual inclination $I$ decreases, the semimajor axis ratio $\alpha$ of the inner pair ($m_1$, $m_2$) with respect to the outer pair ($\mu$, $m_3$) decreases, and the mass ratio of the outer satellite with respect to the inner satellite $\lambda$ increases.

**5 DISCUSSION**

In the Solar system, there are obviously three regions where the triple minor planets have been found: the near-Earth asteroids (NEAs), the main belt asteroids, and the Kuiper Belt objects.

The NEAs binary systems, of which two are triple systems, are small in size and have small separations and are rapidly rotating. Some 15% of these are estimated to be binary systems (Richardson & Walsh 2006; Pravec et al 2006; Margot et al. 2002). From observations nearly 4.9% of these binaries are triple systems (Johnston 2012).

The main belt binary and triple asteroids have much larger mass components and larger separations. Again it is estimated that around 2% are binaries (Richardson & Walsh 2006; Merline et al. 2002). From observations around 6.6% of these binaries are in the form of triple systems (Johnston 2012).

The Kuiper belt binaries are very much larger in mass and have very large separations. Estimates of 5% (Richardson & Walsh 2006) or even more (Stephens & Noll 2006) are in the form of binaries. Again from observations 2.6% of these binaries are at least triple systems (Johnston 2012).

There are several formation hypotheses for the triple minor planets in the Solar system. The triple NEAs are likely to have formed by rotational fission or collision rather than capture (Fang & Margot 2012; Jacobson & Scheeres 2011; Walsh, Richardson & Michel 2012; Marchis et al. 2012; Richardson & Walsh 2006). The triple main belt asteroids are likely to have formed by collision rather than capture or fission (Marchis et al. 2005; Fang et al. 2012; Vokrouhlický et al. 2010; Descamps et al. 2011; Richardson & Walsh 2006). The triple Kuiper belt objects are likely to have formed by gravitational capture (Benecchi et al. 2010), or collision possibly followed by exchange (Ragozzine & Brown 2009), or gravitational collapse (Nesvorný, Youdin & Richardson 2010), or even rotational fission (Ortiz et al. 2012). Note that the results in this paper do not conflict with the collisional formation hypothesis of the triple systems 87 Sylvia (Marchis et al. 2005; Vokrouhlický et al. 2010; Fang et al. 2012), 216 Kleopatra (Descamps et al. 2011) and 136108 Haumea (Ragozzine & Brown 2009). These triple minor planets could experience collisions a very long time ago, and arrived at their present Hill stable state.

Since the triple minor planets discussed were, bar one (3749 Balam for which the incomplete orbital parameters make the Hill stability of the system uncertain), found to be stable, this suggests that there could be more such stable triple minor planets yet to be observed.

There is observational evidence of resolved dust discs around many solar type stars, at both large and small radii indicating the existence of extrasolar counterparts to the main asteroid belt and Kuiper belt in the Solar system (Moro-Martín et al. 2008). The critical parameters for Hill stability of triple minor planets in the Solar system that we have used are independent of scale and our theory of stability could easily be applied to triple systems with similar architectures present in such extrasolar planetary systems

# 6 CONCLUSIONS

In this paper, the Hill stability of the nine known triple minor planets in the Solar system is examined. For the large size ratio triple minor planets, the sufficient condition for Hill stability can be expressed in closed form. For the triple minor planets that are of comparable size, the extremum conditions of the Hill stability are applied. The critical Hill stability surfaces, by variations of some parameters, are plotted for all the nine known triple minor planets. It is shown that the Hill stability regions increase as the mutual inclination of the inner orbit and the outer orbit decreases, the semimajor axis ratio of the inner orbit with respect to the outer orbit decreases, and the mass ratio of the outer satellite with respect to the inner satellite $\lambda$ increases. It is also found that the known triple minor planets in the Solar system are all Hill stable except 3749 Balam, for which the incomplete orbital parameters make the Hill stability of the system uncertain. We conclude that no mass exchange or collision will happen for all time for the eight Hill stable triple minor planets.

**ACKNOWLEDGMENTS**

We are grateful to the anonymous reviewer for an expeditious review and constructive comments. We thank Franck Marchis (SETI Institute), Pascal Descamps (Institut de Mécanique Céleste et de Calcul des Éphémérides), and Frédéric Vachier (Institut de Mécanique Céleste et de Calcul des Éphémérides) who provide us valuable information about multiple asteroids. We also thank Dr. Darin Ragozzine (ITC Fellow, Harvard Institute for Theory and Computation), Ms. Julia Fang (Department of Physics & Astronomy, University of California, Los Angeles), and Laurène Beauvalet (Institut de Mécanique Céleste et de Calcul des Éphémérides) for providing us valuable information about triple systems and useful references. This work was supported by National Basic Research Program of China (973 Program) (2012CB720000) and the National Natural Science Foundation of China (No. 11072122).